\documentclass[aps,prb,preprint,superscriptaddress,amsmath,amssymb,floatfix]{revtex4-2}
\usepackage{graphicx}
\usepackage{booktabs}
\usepackage{hyperref}

\setcitestyle{super}

\date{\today}

\begin{document}

\title[Spin-Selective Electron Transport Through Single Chiral Molecules]{Spin-Selective Electron Transport Through Single Chiral Molecules}

\author{Mohammad Reza Safari}\email{m.safari@fz-juelich.de}
\affiliation{Peter Grünberg Institute, Electronic Properties (PGI-6), Forschungszentrum Jülich, 52425, Jülich, Germany}
\affiliation{Jülich Aachen Research Alliance (JARA-FIT), Fundamentals of Future Information Technology, Forschungszentrum Jülich, 52425, Jülich, Germany}

\author{Frank Matthes}
\affiliation{Peter Grünberg Institute, Electronic Properties (PGI-6), Forschungszentrum Jülich, 52425, Jülich, Germany}
\affiliation{Jülich Aachen Research Alliance (JARA-FIT), Fundamentals of Future Information Technology, Forschungszentrum Jülich, 52425, Jülich, Germany}

\author{Claus M. Schneider}
\affiliation{Peter Grünberg Institute, Electronic Properties (PGI-6), Forschungszentrum Jülich, 52425, Jülich, Germany}
\affiliation{Jülich Aachen Research Alliance (JARA-FIT), Fundamentals of Future Information Technology, Forschungszentrum Jülich, 52425, Jülich, Germany}
\affiliation{Fakultät für Physik, Universität Duisburg-Essen, 47057, Duisburg, Germany}

\author{Karl-Heinz Ernst}\email{karl-heinz.ernst@empa.ch}
\affiliation{Molecular Surface Science Group, Empa, Swiss Federal Laboratories for Materials Science and Technology, 8600, Dübendorf, Switzerland}
\affiliation{Nanosurf Laboratory, Institute of Physics, The Czech Academy of Sciences, 16200, Prague, Czech Republic}
\affiliation{Institut für Chemie, Universität Zürich, 8057, Zürich, Switzerland}

\author{Daniel E. Bürgler}\email{d.buergler@fz-juelich.de}
\affiliation{Peter Grünberg Institute, Electronic Properties (PGI-6), Forschungszentrum Jülich, 52425, Jülich, Germany}
\affiliation{Jülich Aachen Research Alliance (JARA-FIT), Fundamentals of Future Information Technology, Forschungszentrum Jülich, 52425, Jülich, Germany}

\begin{abstract}
The interplay between chirality and magnetism has
been a source of fascination among scientists for over a century.  In
recent years, chirality-induced spin selectivity (CISS) has attracted
renewed interest.  It has been observed that electron transport through
layers of homochiral molecules leads to a significant spin polarization of
several tens of percent.  Despite the abundant experimental evidence
gathered through mesoscopic transport measurements, the exact mechanism
behind CISS remains elusive.  In this study, we report spin-selective
electron transport through single helical aromatic hydrocarbons that were
sublimed in vacuo onto ferromagnetic cobalt surfaces and examined with
spin-polarized scanning tunneling microscopy (SP-STM) at a temperature of
5\,K. Direct comparison of two enantiomers under otherwise identical
conditions revealed magnetochiral conductance asymmetries of up to 50\%
when either the molecular handedness was exchanged or the magnetization
direction of the STM tip or Co substrate was reversed.  Importantly, our
results rule out electron-phonon coupling and ensemble effects as primary
mechanisms responsible for CISS.
\end{abstract}


\maketitle
\newpage

\section*{Introduction}
A characteristic hallmark of life is the mirror asymmetry at the biomolecular level, the so-called homochirality of its building blocks such as amino acids and sugars. As the origin of this asymmetry of life remains elusive, attempts have been made to relate this observation to fundamental forces in the universe. \cite{Pasteur1860, bonner1995} Because electrons from radioactive $\beta$-decay are spin-polarized due to parity violation of the weak nuclear force, Vester and Ulbricht pointed out that this fact should cause either direct asymmetric interaction of electrons with chiral matter or asymmetric radiolysis by their bremsstrahlung. \cite{ulbricht1962} While Farago and Kessler have indeed observed spin-polarized electron scattering from chiral molecules in the vapor phase, the observed asymmetries were only of the order of $10^{-4}$. \cite{campbell1985,mayer1996} However, Blum and coworkers calculated enantio-differential scattering of spin-polarized electrons of up to 80\% for oriented molecules and proposed experiments with surface-aligned chiral molecules. \cite{busalla2002} More recently, asymmetric crystallization induced by polarized electrons and positrons emanating from radioactive sources have been reported. \cite{mahurin2001}

For ordered helical systems aligned on surfaces, large asymmetries in photo current depending on the sense of circular light polarization have been reported. \cite{Ray1999} Spin-selective detection was then used to associate this so-called chirality-induced spin selectivity (CISS) with the state of electron polarization. \cite{Gohler2011} Opposite spin polarization effects of similar magnitude were also found for molecular monolayers of enantiomers of helical hydrocarbons, so-called helicenes. \cite{Kettner2018} Among the various methods employed previously to study the CISS effect, spin-dependent electron transport measurements through chiral molecules have often been applied. In magnetoresistance devices, the magnetic reference electrode is thereby replaced by a monolayer of homochiral molecules in order to measure the spin-dependent molecular conductance (resistance). \cite{xie2011,mathew2014, kettner2015,mondal2015, bloom2016,Kiran2016,Kiran2017, varade2018, Tassinari2018b,Bullard2019,kulkarni2020,Nguyen2022} In many cases, such devices have been realized with the tips of scanning probe microscopes (STM and AFM), still contacting an ensemble of molecules with the uncertainty of a defined contact. In addition, break junctions have been employed. \cite{Aragones2017,slawig2020,yang2023} However, all contact-based transport studies suffer from poor reproducibility in forming identical single-molecule junctions. Another limitation of these experimental approaches is the inability to study both enantiomers in direct comparison, as separate fabrication of the junctions is required, leading to non-identical experimental conditions. Finally, molecular deposition under ambient conditions leaves the exact adsorption geometry unknown, but it is imperative for theoretical modeling and simulation of the experimental results. 

In this study, the aforementioned shortcomings are circumvented by performing spin-polarized scanning tunneling microscopy and spectroscopy (SP-STM/STS) measurements on single chiral heptahelicene ([7]H) molecules sublimed in vacuo from a racemic mixture onto single-crystalline cobalt bilayer nanoislands on a Cu(111) surface. Submolecular resolution of SP-STM/STS allows examination of spin selectivity of individual enantiomers under well-defined conditions. This single-molecule approach represents a major advance over previous studies, as it allows  to distinguish the handedness of molecules adsorbed on a substrate with a given magnetization. Hence, the conductance of enantiomers can be compared while the electrodes (same Co island and same STM tip) remain unchanged. The direct measurement of the tunneling current through single enantiomer molecules on the same Co nanoisland yields magnetochiral conductance asymmetries of up to 50\%.

\section*{Results and Discussion}

\begin{figure*}[b]
\begin{center}
\includegraphics[width=0.94\textwidth]{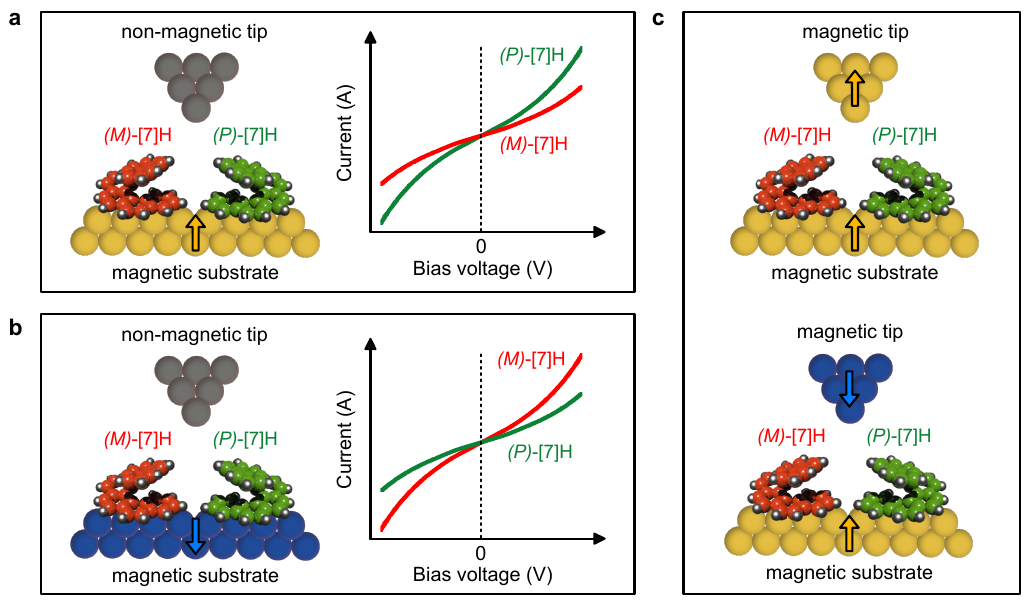}
\caption{Evaluation of magnetochiral electron tunneling asymmetries with SP-STM. (a) On a ferromagnetic surface with out-of-plane magnetization, magnetochiral tunneling will show different conductance for the enantiomers, here \textit{(M)}-[7]H and \textit{(P)}-[7]H. (b) The effect will reverse on a surface with opposite out-of-plane magnetization direction. (c) For STM conductance measurements on a ferromagnetic surface with a magnetic STM tip, the conductance of the enantiomers will switch when the tip magnetization is reversed.}
\label{principles}
\end{center}
\end{figure*}

Magnetochiral asymmetries in electron tunneling are expected for example for enantiomers on a surface with uniform out-of-plane magnetization. Figure~\ref{principles} presents principle approaches for identifying magnetochiral asymmetries with SP-STM/STS. Using a non-magnetic STM tip, enantiomers will show different conductances on an out-of-plane magnetized ferromagnetic surface (Figure~\ref{principles}a). In order to quantify differences between the $I-V$ spectra of different enantiomers, the so-called magnetochiral conductance asymmetry (MChA) is here defined as
\begin{equation}
    \text{MChA}=\frac{ I_\mathrm{(P)}-I_\mathrm{(M)} }{ I_\mathrm{(P)}+I_\mathrm{(M)} },
\label{MChA}    
\end{equation}
where $I_\mathrm{(P)}$ and $I_\mathrm{(M)}$ represent the tunneling current passing through the \textit{(P)}- and \textit{(M)}-[7]H molecule, respectively. If the measurement is repeated on a surface with opposite out-of-plane magnetization direction, the $I-V$ spectra of the enantiomers will interchange (Figure~\ref{principles}b). In order to quantify the differences between the $I-V$ spectra measured on oppositely magnetized surfaces, the enantiospecific magnetic conductance asymmetry (EMA) is here defined as
\begin{equation}
    \text{EMA}=\frac{ I_\mathrm{\text{up}}-I_\mathrm{\text{down} }}{ I_\mathrm{\text{up}}+I_\mathrm{\text{down}} },
\label{EMA}    
\end{equation}
where $I_\mathrm{\text{up}}$ and $I_\mathrm{{down}}$ correspond to the measured tunneling current through molecules of identical absolute handedness but adsorbed on surfaces with opposite out-of-plane magnetization.
Another way of identifying magnetochiral asymmetries is using oppositely magnetized ferromagnetic STM tips plus evaluating $I-V$ spectra of enantiomeric pairs on a surface with fixed magnetization direction (Figure~\ref{principles}c). This method does not require measurements on different surfaces and avoids averaging of $I-V$ curves of different molecules. Again, the two enantiomers on a single magnetic domain should show different conductances solely based on exchanging the absolute handedness of adsorbed molecules.

An ideal ferromagnetic surface system for such studies is the Co/Cu(111) surface. Triangular Co bilayer nanoislands are formed by in-situ sublimation of a sub-monolayer of Co onto a pre-cleaned (111) surface of a Cu single crystal. The formation process, atomic structure, electronic properties, and magnetic characteristics are well established. \cite{DeLaFiguera1993, Pietzsch2004, pietzsch2006, Negulyaev2008, oka2010} It has been previously reported that the controlled deposition of organic molecules on the surface of these highly reactive ferromagnetic nanoislands does not impair the magnetization of the underlying substrate. \cite{Esat2017,Safari2022,safari2022enantio} By using a Co-functionalized tip, the different out-of-plane magnetization directions of the Co-nanoislands (i.e., pointing into vacuum or substrate) can be determined.

\begin{figure*}[b]
\begin{center}
\includegraphics[width=0.83\textwidth]{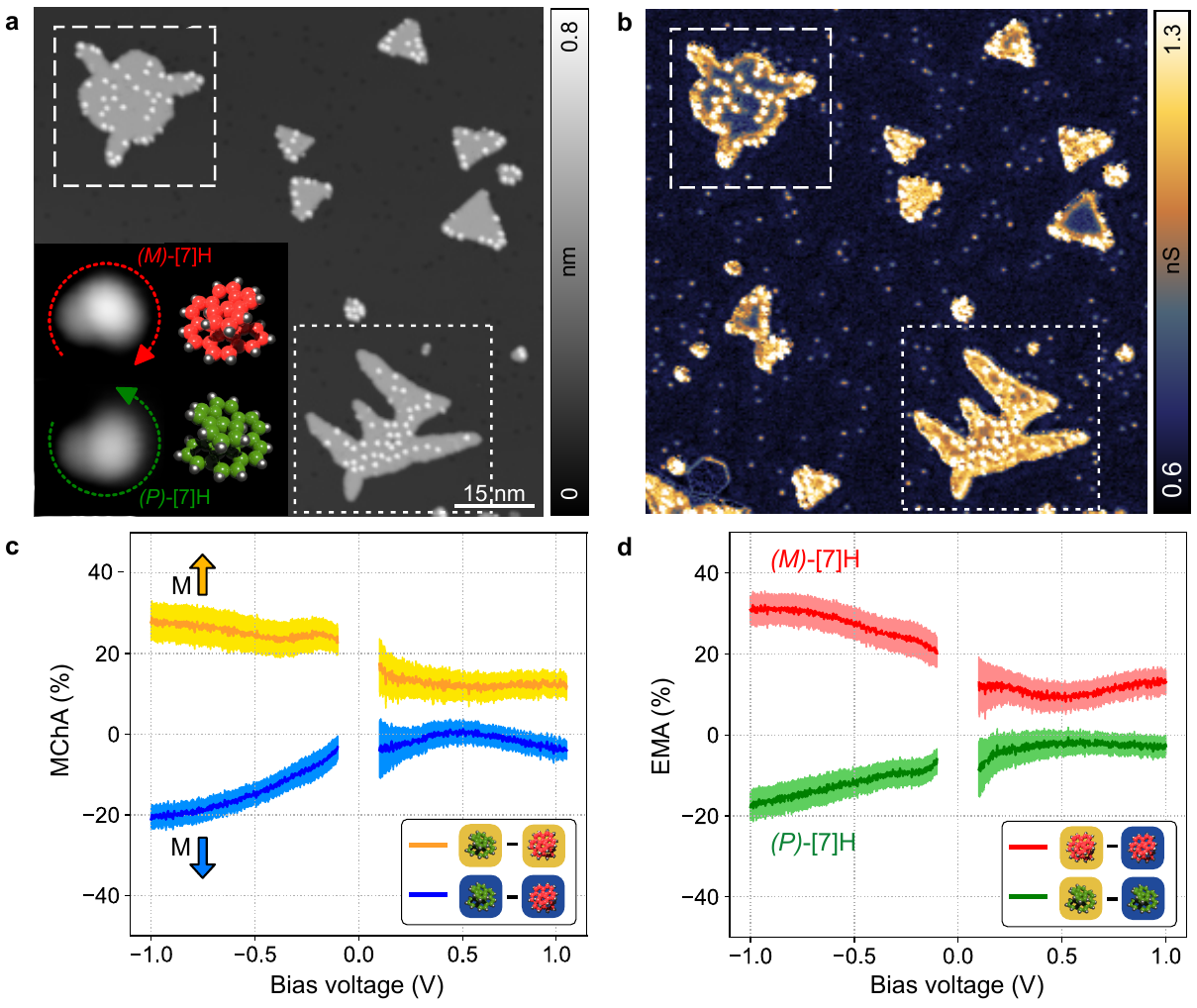}
\caption{Magnetochiral tunneling conductance asymmetries.
(a) Constant-current topographic STM image showing adsorbed [7]H molecules that are exclusively attached to the Co nanoislands. The inset shows high-resolution topographic images of the enantiomers. The determined handedness of the \textit{(P)}- and \textit{(M)}-[7]H enantiomer is indicated by a green and red circular arrow, respectively. (b) The $dI/dV$ map recorded simultaneously with (a) reveals magnetic contrast between Co nanoislands with opposite out-of-plane magnetization directions, here displayed by blue and yellowish false colors ($V_\mathrm{bias}=-600$\,mV, $I_\mathrm{t}=550$\,pA, $V_\mathrm{mod}=20$\,mV, $f_\mathrm{mod}=752$\,Hz, 5\,K, Co-functionalized W tip).
(c) and (d) Magnetochiral asymmetry MChA and enantiospecific magnetic asymmetry EMA curves of [7]H enantiomers calculated according to Eqs.~\ref{MChA} and \ref{EMA}, respectively, from $I-V$ spectra measured with a non-magnetic tip. The dark curves display the calculated values, and the light colored areas represent the standard error resulting from averaging the $I-V$ spectra of five molecules.
The values near $V_\mathrm{bias}=0$ diverge due to division by small values and have been masked out.}
\label{statistics}
\end{center}
\end{figure*}

Figure~\ref{statistics}a shows a constant-current topographic STM image of [7]H chiral molecules taken with a Co-functionalized STM tip (see Supporting Information S1). The molecules were sublimed at 400\,K from a racemic mixture under ultra-high vacuum (UHV) conditions onto a Cu(111) crystal covered by about 10\% with previously formed Co nanoislands. During [7]H deposition, the substrate was at room temperature (RT), and the deposited amount of [7]H molecules corresponds to less than 2\% of a close-packed monolayer (ML) on Cu(111). The Co nanoislands are decorated with [7]H molecules, while the Cu surface in between the islands remains bare. This is due to high surface mobility of the molecules upon adsorption at RT and the stronger binding of [7]H to Co than to Cu. \cite{Safari2022,safari2022enantio}

Figure~\ref{statistics}b shows the differential conductance ($dI/dV$) map recorded simultaneously with the topographic image in Figure~\ref{statistics}a (see Supporting Information S1). The different $dI/dV$ signal of the islands arises from opposite magnetization directions of the out-of-plane magnetized islands \cite{pietzsch2006,oka2010} and hence confirms that the islands are ferromagnetic even in the presence of chemisorbed [7]H molecules. \cite{Esat2017,safari2022enantio,metz21-00} The dashed and dotted framed areas in Figures~\ref{statistics}a and b indicate two Co islands that are oppositely magnetized. High-resolution constant-current topography images of each selected island reveal the chirality of the adsorbed molecules. The inset in Figure~\ref{statistics}a illustrates an example of two adsorbed enantiomers on the island with dark $dI/dV$ contrast. The resolved submolecular structure can be attributed to the topographic profile of the [7]H molecules, which adsorb with their helical axis basically perpendicular to the surface. The handedness of the molecules is reflected in the sense of rotation of the screw-like increasing height variations, as indicated by red and green circular arrows for \textit{(M)}-[7]H and \textit{(P)}-[7]H enantiomers, respectively. \cite{Safari2022,safari2022enantio}

After identification of islands with opposite out-of-plane magnetization with a Co-functionalized tip, the tip was re-modified with copper such that no longer any magnetic contrast has been observed (see Supporting Information S2). With this non-magnetic tip, current-voltage ($I-V$) spectra from ten molecules (five \textit{(P)}- and five \textit{(M)}-[7]H enantiomers) on each of the framed islands in Figure~\ref{statistics}a were measured (for measurement procedure see Supporting Information S1). $I-V$ spectra of single molecules and averaged spectra are shown in Supporting Information S3. The magnetochiral conductance asymmetries (MChA, as defined in Eq.~\ref{MChA}) between enantiomers on identical islands (i.e., with identical magnetization) is plotted in Figure~\ref{statistics}c as a function of applied bias voltage. The enantiospecific magnetic conductance asymmetries (EMA, as defined in Eq.~\ref{EMA}) for \textit{(P)}- and \textit{(M)}-[7]H molecules on oppositely magnetized Co island (i.e., islands with bright and dark magnetic $dI/dV$ contrast) are plotted in Figure~\ref{statistics}d. For both evaluation methods clear asymmetries are observed. Both pairs of curves are significantly separated from each other by more than the standard errors, which clearly confirms the CISS effect at the single-molecule level. Previous conclusions based on ensemble effects at work in CISS are therefore questionable. \cite{Nguyen2022,dianat2020} Interestingly, the qualitative similarities between MChA and EMA curves suggest that exchanging the chirality of the molecules (Figure~\ref{statistics}c) or reversing the magnetization of the substrate (Figure~\ref{statistics}d) produce similar effects on the conductance measurements.

\begin{figure*}[t]
\begin{center}
\includegraphics[width=0.81\textwidth]{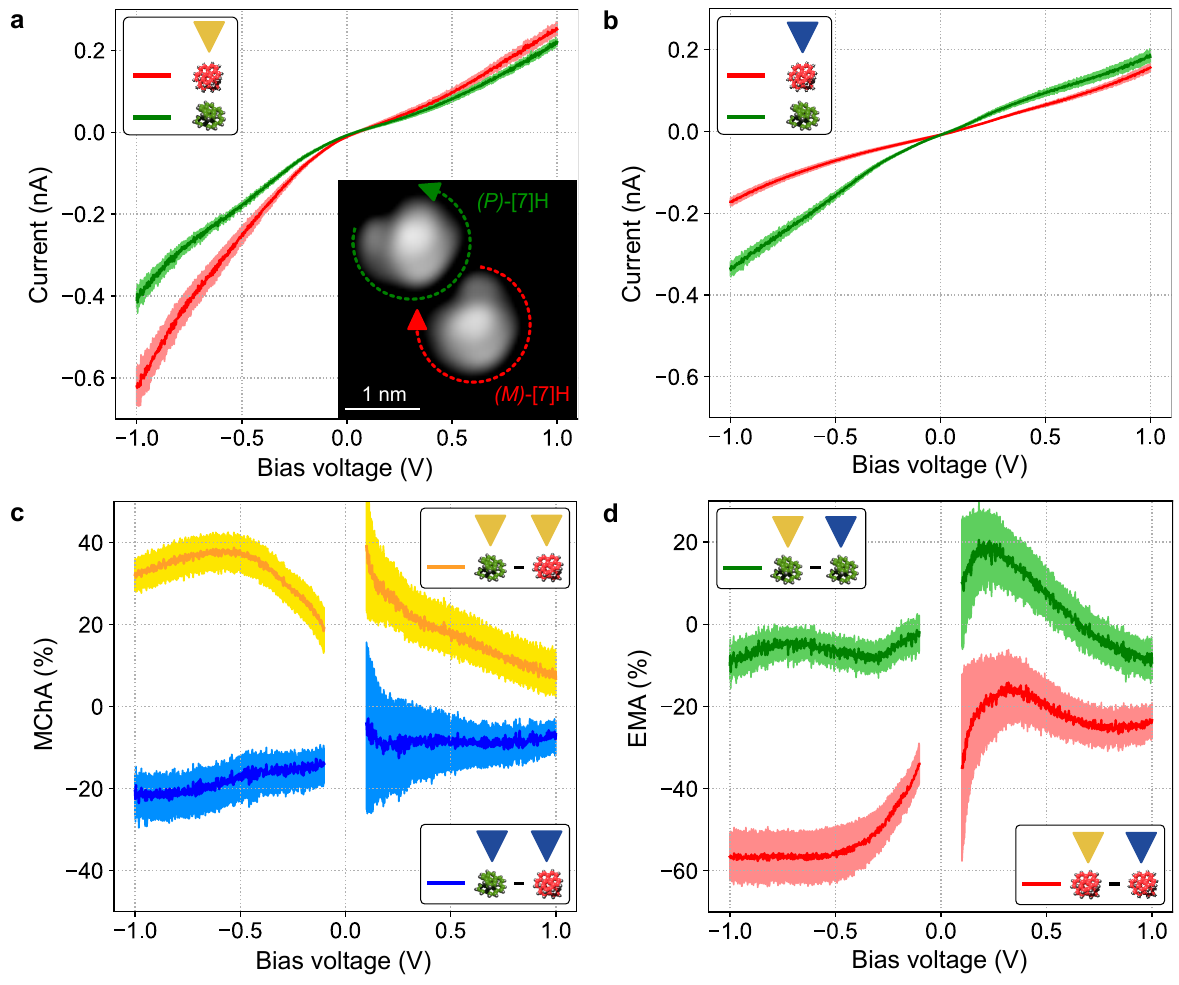}
\caption{Magnetochiral asymmetries with a magnetic STM tip. (a) $I-V$ spectra acquired for both [7]H enantiomers using a first magnetization direction of the STM tip ($V_\mathrm{stab}=2$\,V and $I_\mathrm{stab}=800$\,pA). The inset shows a high-resolution topographic image of the [7]H enantiomer pair with the determined handedness of the molecules indicated by green and red circular arrows, respectively ($V_\mathrm{bias}=100$\,mV, $I_\mathrm{t}=100$\,pA, 5\,K, Co-functionalized W tip). (b) Same as (a), but utilizing a second magnetic tip configuration with reversed magnetization direction. 
(c) and (d) Magnetochiral asymmetry (MChA) and magnetic asymmetry (EMA) curves of the [7]H enantiomers according to Eqs.\,\ref{MChA} and \ref{EMA}, respectively. 
The dark curves display the calculated values in percent, and the light colored areas represent the propagated error resulting from the standard errors of the two involved single-molecule spectra.}
\label{single-island}
\end{center}
\end{figure*}

To avoid the required averaging of $I-V$ curves over different molecules and in order to reveal true single-molecular properties, $I-V$ spectra of a single pair of enantiomers adsorbed on a single Co nanoisland with a fixed magnetization direction have been acquired. Figure~\ref{single-island}a shows the $I-V$ curves for two enantiomers recorded with a magnetized tip. The dark red and green curves represent here the spectrum of one enantiomer each, and the light colored areas indicate the standard error of the single-molecule spectra (see Supporting Information S1). Under these conditions, the \textit{(M)}-[7]H enantiomer (red curve) shows the larger conductance. In particular for negative bias voltages, meaning that electrons tunnel from the occupied states of the sample into the unoccupied states of the tip, a distinct difference is observed. Such $I-V$ asymmetry for the enantiomers switches if the tip magnetization is reversed (Figure~\ref{single-island}b). The reversal of the tip magnetization upon re-functionalization above a remote Co island is verified by the $dI/dV$ maps presented in Supporting Information S4, which show opposite magnetic contrast.

The respective magnetochiral asymmetry MChA and enantiospecific magnetic asymmetry EMA curves evaluated from the $I-V$ spectra are displayed in Figure~\ref{single-island}c and d. Here, $I_{\text{up}}$ and $I_{\text{down}}$ in Eq.~\ref{EMA} refer to the current measured for the different directions of the tip magnetization of the first and second tip configuration, respectively. The different shapes and magnitudes of the MChA and EMA curves are most likely due to structural and electronic differences of the STM tip induced by the re-functionalization. Such changes strongly affect the EMA curves, for which $I-V$ spectra taken with different tip configurations are subtracted from each other (Eq.~\ref{EMA}). The MChA curves, on the contrary, are each calculated from two single-molecule spectra taken with the same tip configuration (Eq.~\ref{MChA}), so the structural and electronic changes of the tip do not play a role. Remarkably, both pairs of curves are separated by more than the standard errors, clearly demonstrating CISS as a property of a single molecule. As expected for CISS, experiments in which either surface (Figure \ref{statistics}) or tip (Figure \ref{single-island}) magnetization was reversed lead to an effect similar to that of changing molecular handedness.

\begin{figure*}[b]
\begin{center}
\includegraphics[width=1.0\textwidth]{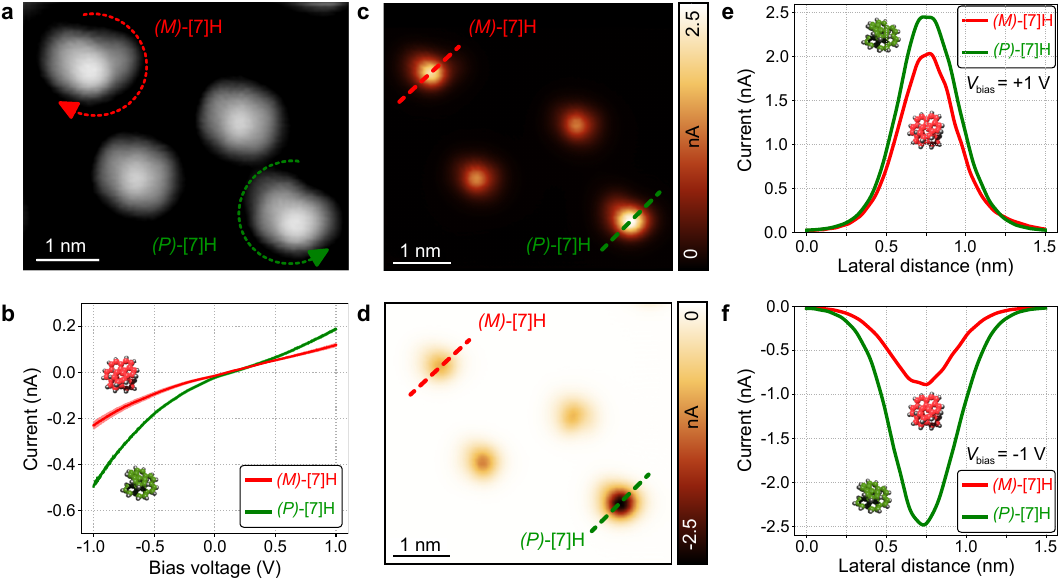}
\caption{CISS for a single pair of [7]H enantiomers.
(a) High-resolution topographic image of a pair of enantiomers with the handedness indicated by circular arrows. The handedness of the two molecules in the middle could not be determined ($V_\mathrm{bias}=-1$\,V, $I_\mathrm{t}=1$\,nA, 5\,K).
(b) $I-V$ spectra of two the [7]H enantiomers in (a) ($V_\mathrm{stab}=2$\,V and $I_\mathrm{stab}=500$\,pA). 
 The light green and red colored areas indicate the standard error of the single-molecule spectra.
(c) and (d) Constant-height STM images (5\,K, Co-functionalized W tip) acquired in the same area as shown in (a) with $V_\mathrm{bias}=1$\,V ($\Delta z_\mathrm{0}=0.20$\,nm) and $V_\mathrm{bias}=-1$\,V ($\Delta z_\mathrm{0}=0.15$\,nm), respectively. Dashed lines indicate the positions of the line profiles across the two enantiomers shown in (e) and (f). 
}
\label{constantH}
\end{center}
\end{figure*}

A final proof of single-molecule magnetochiral asymmetry comes from an experimental procedure in which the constant-height STM mode of operation (see Supporting Information S1) is used to map current variations directly across a single pair of enantiomers, while all other conditions (tip structure, tip height, tip and island magnetization directions, etc.) are kept identical. A topographic constant-current STM image of the pair of enantiomers adsorbed on a Co nanoisland is shown in Figure~\ref{constantH}a. The absolute handedness of the molecules, as identified by their intramolecular topographic STM contrast profiles, is indicated. The corresponding $I-V$ single-molecule spectra are displayed in Figure~\ref{constantH}b. For the given magnetization direction of this Co island, the \textit{(P)}-[7]H enantiomer exhibits a larger conductance than the \textit{(M)}-[7]H enantiomer under such identical experimental conditions. Figures~\ref{constantH}c and d show constant-height STM images at +1\,V and -1\,V bias voltage, respectively, with the current intensity displayed by brightness. The STM tip is scanned across the sample with deactivated feedback loop at a constant distance from the Co surface, and the tunneling current is recorded as a function of lateral position (see Supporting Information S1). From the brightness contrast for both enantiomers in both images, it becomes clear that the enantiomers exhibit different conductance, regardless of the exact lateral tip position on the molecules. In order to quantitatively compare the current through the two enantiomers, line profiles across the molecules (along red and green dashed lines in Figures~\ref{constantH}c and d) are plotted in Figures~\ref{constantH}e and f. The larger amplitudes of the tunneling currents above the \textit{(P)}-[7]H give unequivocal evidence that this enantiomer exhibits a larger conductance than the \textit{(M)}-[7]H enantiomer. The MChA (Eq.~\ref{MChA}) in the current peaks in Figures~\ref{constantH}e and f amounts to 9\% and 48\%, respectively. 

The presented experimental results are based on individual enantiomers that are laterally separated by a few nanometers. The observed magnetochiral asymmetries are thus unequivocal evidence that CISS is a single-molecule property. Previously proposed mechanisms based on intermolecular interactions \cite{dianat2020} or cooperative effects \cite{Nguyen2022} are not needed for explaining CISS effects. All $I-V$ measurements are carried out in the tunneling mode with a vacuum gap between tip and  molecule. Thus, in contrast to contact-based transport studies, uncertainties of contact formation and molecular deformation due to force interaction with the tip do not play a role.

We emphasize that here the CISS effect was observed at 5\,K. The asymmetry values of up to about 50\% are similar those measured at RT for cationic [4]helicenes on HOPG by conducting-probe AFM and using a magnetic Ni tip. \cite{Kiran2016} In contrast, only 6 to 8\% spin polarization of photoelectrons was observed after transmission through a ML of enantiopure [7]H molecules on Cu(332), Ag(110), and Au(111). \cite{Kettner2018} However, these experiments employed photoelectrons travelling 1.18\,eV above the vacuum level, whereas the $I-V$ measurements presented here cover electronic states within $\pm 1$\,eV around the Fermi energy. Despite the challenge in comparing conductance asymmetry and polarization values of different molecule-substrate systems and complementary measurement techniques, our low-temperature data suggest that thermal enhancement mechanisms \cite{das2022} are not essential ingredients for CISS.

\section*{Conclusion}
\sloppy
Sizable CISS effects in electron transport through helical hydrocarbon molecules chemisorbed to ferromagnetic Co nanoislands have been observed microscopically under well-defined conditions in ultrahigh vacuum at the single-molecule level by spin-polarized STM. The minimum setup for detecting these CISS effects comprises a single pair of enantiomers adsorbed on an out-of-plane magnetized Co nanoisland and imaged quasi-simultaneously with the same STM tip. Even at 5\,K, magnetochiral asymmetries in electrical conductance are of the order of 50\%.  Therefore, coupling of electrons to phonons (thermal vibrations) and ensemble effects due to intermolecular interaction and can be ruled out as primary mechanisms of CISS. The well-defined experimental conditions including the specific adsorption geometry and weak coupling to the STM tip render the presented experiments amenable to theoretical modeling and simulation.


%

\end{document}